\documentclass[12pt]{iopart}
\usepackage{graphicx}% Include figure files
\usepackage{epsfig}% Include figure files
\usepackage{bm}% bold math

\begin{document}

\title{Low-energy Dipole Excitations in Nuclei at the $N=50,82$
and $Z=50$ Shell Closures as Signatures for a Neutron Skin. }

\author{N. Tsoneva$^{1,2}$, H. Lenske$^1$ }

\address{$^1$  Institut f\"ur Theoretische Physik, Universit\"at Gie\ss en, Heinrich-Buff-Ring 16, D-35392 Gie\ss en, Germany.}

\address{$^2$  Institute for Nuclear Research and Nuclear Energy, Tzarigradsko chaussee 72, BG-1784  Sofia, Bulgaria }

\ead{tsoneva@theo.physik.uni-giessen.de}

\begin{abstract}
Low-energy dipole excitations have been investigated theoretically
in N=50 $^{88}$Sr and $^{90}$Zr, several N=82 isotones and the
$Z=50$ Sn isotopes. For this purpose a method incorporating both HFB
and multi-phonon QPM theory is applied. A concentration of
one-phonon dipole strength located below the neutron emission
threshold has been calculated in these nuclei. The analysis of the
corresponding neutron and proton dipole transition densities allows
to assign a genuine pattern to the low-energy excitations and making
them distinct from the conventional GDR modes. Analyzing also the
QRPA wave functions of the states we can identify these excitations
as Pygmy Dipole Resonance (PDR) modes, recently studied also in Sn
and N=82 nuclei. The results for $N=50$ are exploratory for an
experimental project designed for the bremsstrahlung facility at the
ELBE accelerator.

\end{abstract}

%\vspace{2pc}\multimap%\noindent{\it Keywords}: Article preparation, IOP journals

\maketitle

\section{Introduction}
The progress in nuclear structure physics is closely connected with
the extended experimental possibilities of the new facilities
\cite{PDRrev:06}. In particular the experiments with rare isotope
beams give us the opportunity to investigate nuclei far from
stability. One of the most interesting results was the discovery of
a new dipole mode at low-excitation energy in nuclei with high
isospin asymmetry \cite{PDRrev:06,Rye:02,Adrich:2005}. Typically,
one observes in nuclei with a neutron excess, $N>Z$, a concentration
of electric dipole strength of predominantly electric character at
or close to the particle emission threshold. Since this bunching of
$1^-$ states resembles spectral structures, otherwise known to
indicate resonance phenomena, these states have been named Pygmy
Dipole Resonance (PDR). However, only a tiny fraction, less than 1\%
of the total Thomas-Reiche-Kuhn energy weighted dipole sum rule
strength is found in the PDR region. These studies have various
astrophysical applications, e.g. the explosive nucleosynthesis and
the neutron stars.

Here, we present our investigations on the dipole excitations in
many nuclei from N=50, 82 and Z=50 regions. For this purpose a
method based on Hartree-Fock-Bogoljubov (HFB) description of the ground state is applied
\cite{Ts04}. The excited states are calculated with the Quasiparticle-Phonon Model (QPM)
\cite{Sol76}.

\section{The Model}

The model Hamiltonian \cite{Sol76}:
\begin{equation}
{H=H_{MF}+H_M^{ph}+H_{SM}^{ph}+H_M^{pp}} \quad
\label{hh}
\end{equation}
is built from the HFB term $H_{MF}=H_{sp}+H_{pair}$ containing two
parts: $H_{sp}$ describes the motion of protons and neutrons in a
static, spherically-symmetric mean-field, taken as a Wood-Saxon (WS)
potential. The parameters of the WS potential are derived from fully
microscopic HFB calculations of the ground state
\cite{Hof98,Ts04,Ts07}, separately for every nucleus under
consideration, which is different from the standard QPM scheme given
in \cite{Sol76}; $H_{pair}$ accounts for the monopole pairing
between isospin identical particles with coupling constants
extracted from the data \cite{Audi95}. The last three terms present
the residual interaction $H_{res}=H_M^{ph}+H_{SM}^{ph}+H_M^{pp}$ and
refer to the multipole-multipole $H_M^{ph}$ and spin-multipole
interactions $H_{SM}^{ph}$ of isoscalar and isovector type in the
particle-hole  and multipole pairing $H_M^{pp}$ in the
particle-particle channels.

In the QPM the residual interaction is taken in a separable form
\begin{equation}
\Re_{\lambda}(r_{1},r_{2})=\kappa^{\lambda}\Re_{\lambda}
(r_{1})\Re_{\lambda}(r_{2}) ,
\end{equation}
where $\Re_{\lambda}(r)$ is a radial form factor , which is usually
chosen as $r^{\lambda}$; $
\kappa^{\lambda}=(\kappa^{\lambda}ннн_{0},\kappa^{\lambda}ннн_{1})$
are empirical isoscalar and isovector coupling constants, which are
obtained by a fitting procedure \cite{Vdo83}.

The nuclear excited states are constructed of
Quasiparticle-Random-Phase-Approximation (QRPA) phonons, defined as
a linear combination of two-quasiparticle creation and annihilation
operators as follows:
\begin{equation}
Q^{+}_{\lambda \mu i}=\frac{1}{2}\sum_{\tau}^{n,p}{
\sum_{jj'}{ \left(\psi_{jj'}^{\lambda i}A^+_{\lambda\mu}(jj'\tau)
-\varphi_{jj'}^{\lambda i}\widetilde{A}_{\lambda\mu}(jj'\tau)
\right)}},
\label{eq:StateOp}
\end{equation}
where ${A}^+_{\lambda \mu}$ and $\widetilde{A}_{\lambda \mu}$ are
time-forward and time-backward\footnote{The time reversed operator
is defined as $\widetilde{A}_{\lambda \mu}=(-)^{\lambda
-\mu}A_{\lambda-\mu}$} operators, coupling proton and neutron
two-quasiparticle creation or annihilation operators to a total
angular momentum $\lambda$ with projection $\mu$ by means of the
Clebsch-Gordan coefficients $C^{\lambda\mu}_{jmj'm'}=\left\langle
jmj'm'|\lambda\mu\right\rangle$. Correspondingly,

\begin{equation}
A^+_{\lambda \mu}(j_1j_2q)=\left[\alpha^+_{j_1q}\alpha^+_{j_2q}\right]_{\lambda\mu}=
\sum_{m_1m_2}C^{\lambda\mu}_{j_1m_1j_2m_2}\alpha^{+}_{j_1} \alpha^{+}_{j_2}
\end{equation}

The QRPA phonon operators obey the equation of motion
\begin{equation}\label{eq:EoM}
\left[H,Q^+_\alpha\right]=E_\alpha Q^+_\alpha \quad ,
\end{equation}
which solves the eigenvalue problem, giving the excitation energies
$E_\alpha$ and the wave functions of the excited states, defined by
the time-forward and time-backward amplitudes $\psi_{jj'}^{\lambda
i}$ and $\varphi_{jj'}^{\lambda i}$), respectively.

The spacial structure of a nuclear excitation becomes accessible by
analyzing the one-body transition densities $\delta\rho (\vec{r})$,
which are the non-diagonal elements of the nuclear one-body density
matrix. In the QRPA theory the one-phonon transition density is
given by the coherent sum over two-quasiparticle transition
densities entering in the wave function of a phonon by the relation:
\begin{equation}
\rho_{\lambda i} (r)=\sum_{j_1\geq j_2}{ \rho_{j_1j_2}^{(\lambda)}
(r)g^{\lambda i}_{j_1j_2}} \quad , \label{phoph}
\end{equation}
where the radial parts are given by the radial single particle wave
functions and reduced matrix elements
\begin{equation}
\rho^\lambda_{j_1j_2q}(r)=R^*_{j_1q}(r)R_{j_2q}(r)\frac{1}{\hat{\lambda}}\langle
j_1||i^\lambda Y_\lambda||j_2\rangle \quad ,
\end{equation}
with $\hat{\lambda}=\sqrt{2\lambda+1}$.
The BCS quasiparticle properties and QRPA state amplitudes are
contained in
\begin{equation}
g^{\lambda i}_{j_1j_2}=\frac{\psi_{j_1j_2}^{\lambda
i}+\varphi_{j_1j_2}^{\lambda i}}{1+\delta_{j_1j_2}}\left(u_{j_1}
v_{j_2}+u_{j_2}v_{j_1}\right) \quad .
\end{equation}

\section{Application to PDR Excitations}
\begin{center}
\begin{figure}
\includegraphics[width=9cm,height=8.5cm,angle=0]{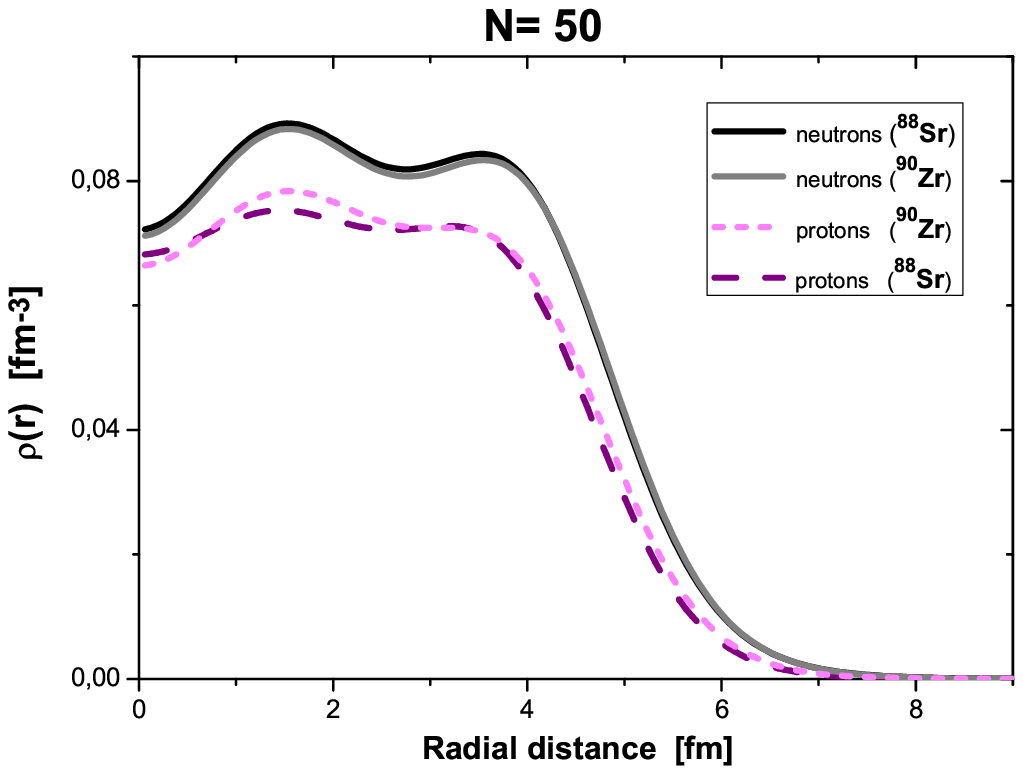}
\includegraphics[width=9cm,height=9cm,angle=0]{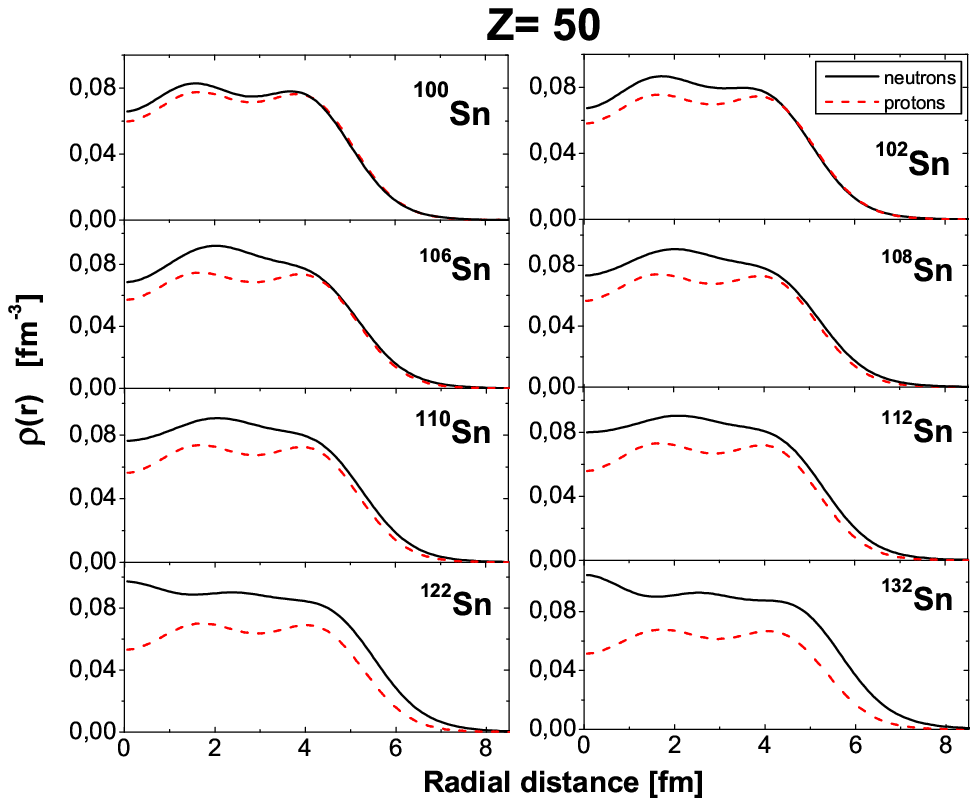}
\caption{Ground state densities of N=50 and Z=50 isotopes used in
the QPM calculations.} \label{fig1}
\end{figure}
\end{center}

\begin{figure}
\includegraphics[width=11cm,height=9cm,angle=0]{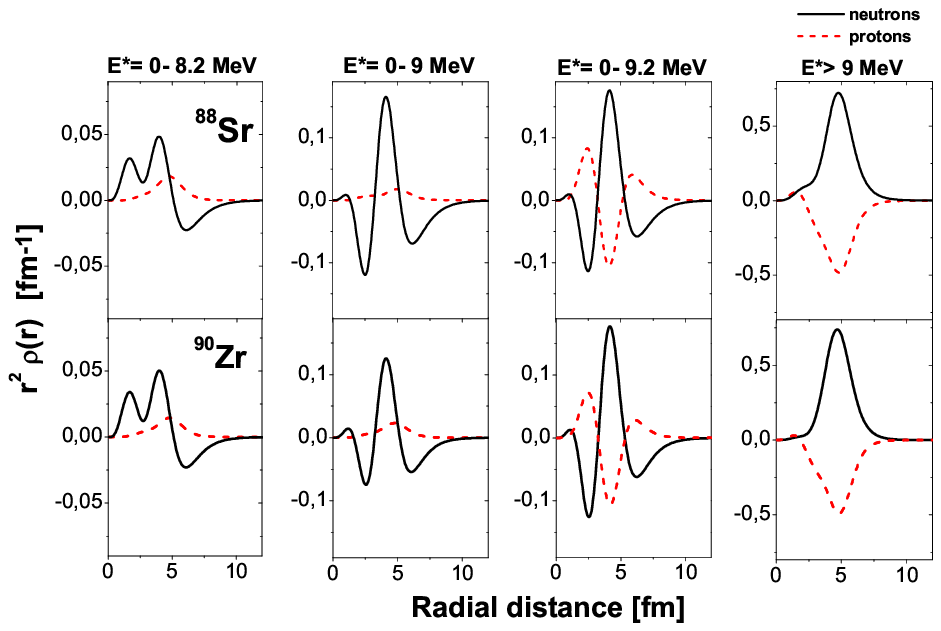}
\includegraphics[width=7cm,height=8cm,angle=0]{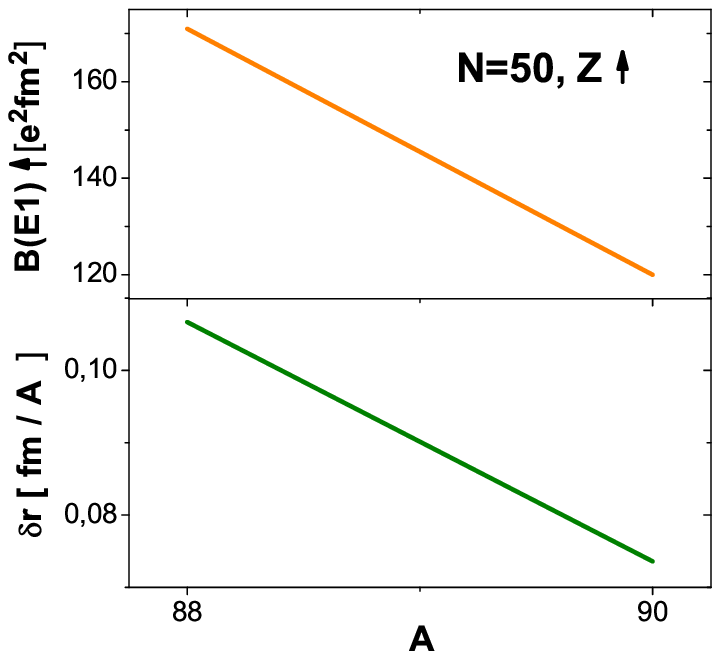}
\caption{Dipole one-phonon transition densities in N=50 nuclei
(left). The total PDR strength is compared to the nuclear skin
thickness $\delta r$, eq.\protect\ref{eq:skin}, as a function of the
mass number in N=50 nuclei (right).} \label{fig2}
\end{figure}

\begin{figure}
\includegraphics[width=11cm,height=10cm,angle=0]{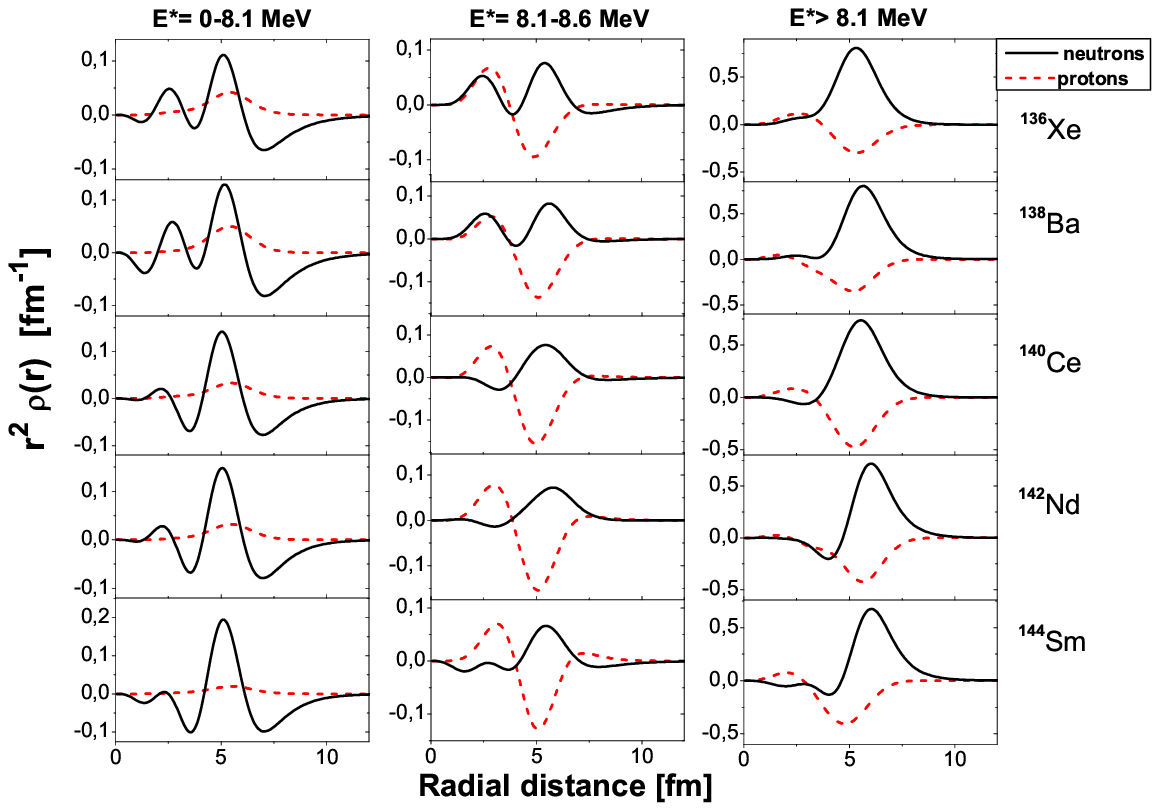}
\includegraphics[width=7cm,height=9cm,angle=0]{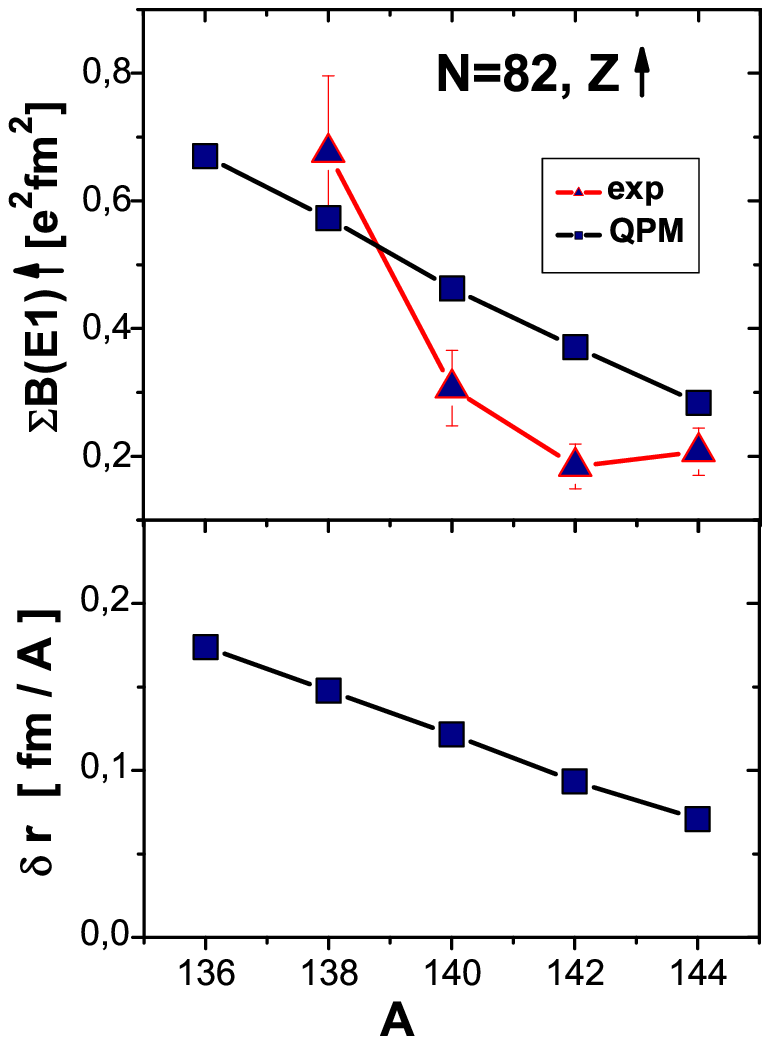}
\caption{One-phonon dipole transition densities in N=82 nuclei
(left). The total PDR strength is compared to the nuclear skin
thickness $\delta r$, eq.\protect\ref{eq:skin}, as a function of the
mass number in N=82 nuclei (right).} \label{fig3}
\end{figure}

\begin{figure}
\includegraphics[width=9cm,height=8cm,angle=0]{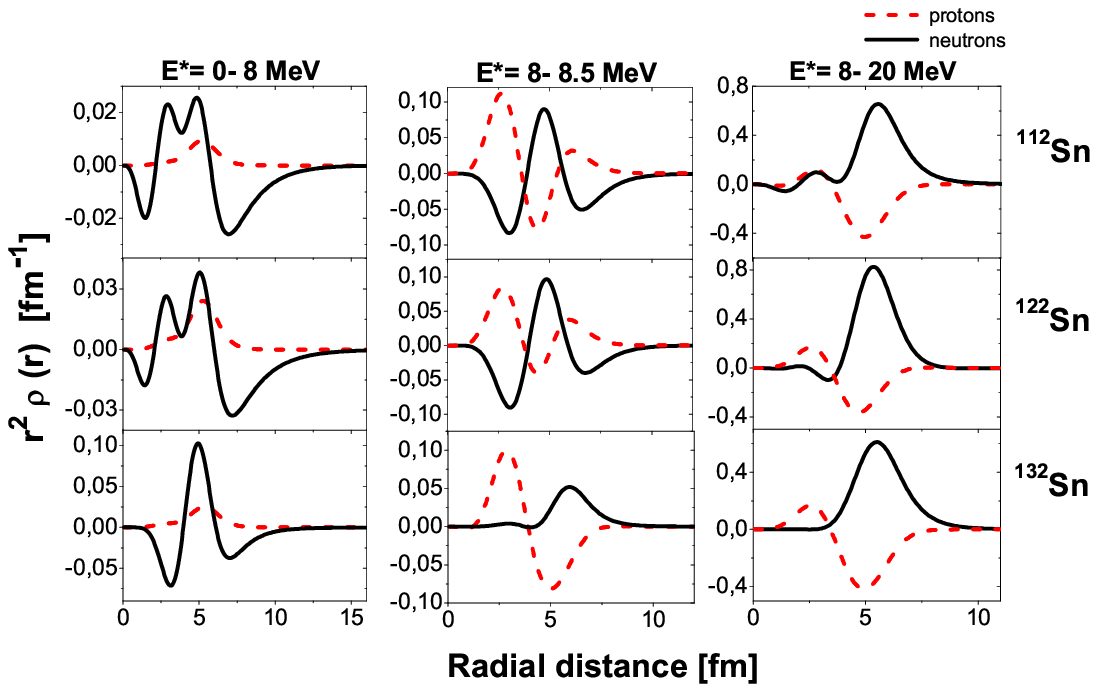}
\includegraphics[width=9cm,height=9cm,angle=0]{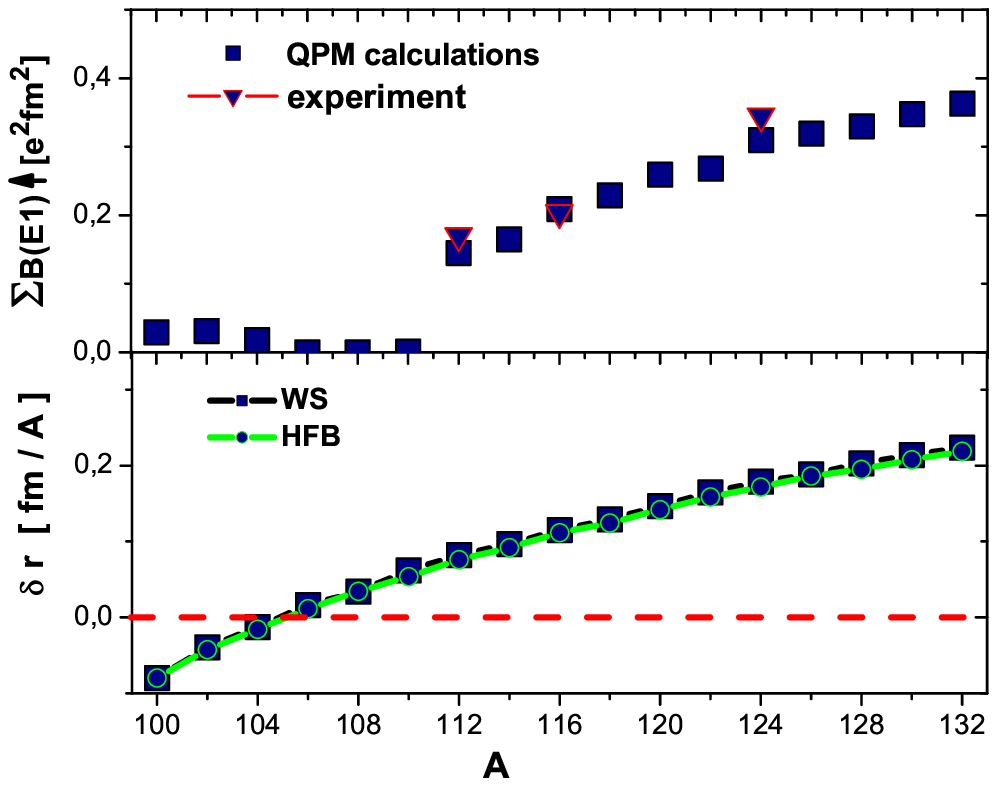}
\caption{Dipole one-phonon transition densities in Z=50 nuclei
(left). The total PDR strength is compared to the nuclear skin
thickness $\delta r$, eq.\protect\ref{eq:skin}, as a function of the
mass number in Z=50 nuclei (right).} \label{fig4}
\end{figure}

\begin{figure}
\includegraphics[width=10cm,height=5cm,angle=0]{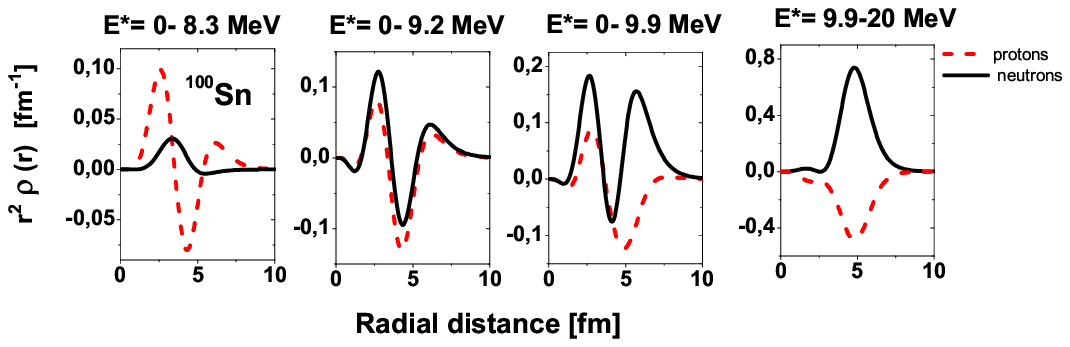}
\caption{QPM results for the one-phonon dipole transition densities
in $^{100}$Sn.} \label{fig5}
\end{figure}

The calculated neutron and proton ground state densities are
presented in Fig.\ref{fig1} for N=50 and Z=50 nuclei. Of special
importance for our investigations are the surface regions, where the
formation of a skin takes place. For the N=50 nuclei, the neutron
skin decreases from $^{88}$Sr to $^{90}$Zr, when the number of the
protons increases. In the $Z=50$ Sn isotopes we find, that for
A$\geq$106 the neutron distributions begin to extend beyond the
proton density and the effect continues to increase with the neutron
excess, up to $^{132}Sn$. Thus, these nuclei have a neutron skin.
The situation reverses in $^{100-102}$Sn, where a tiny proton skin
appears at the nuclear surface. We find, that the properties of the
ground states reflect directly off the low-energy dipole
excitations. From QRPA calculations in N=50, N=82 \cite{Volz06} and
$^{112-132}$Sn \cite{Ts04,Ts07} nuclei a sequence of  low-lying
one-phonon dipole states of almost pure neutron structure, located
below the particle threshold is obtained. The analysis of the dipole
transition densities at E*$\le$8 MeV in Z=50 (Fig.\ref{fig4} left),
N=82 (Fig.\ref{fig3} left) and at E*$\le$9 MeV in N=50
(Fig.\ref{fig2} left) reveal in-phase oscillation of protons and
neutrons in the nuclear interior, while at the surface only neutrons
contribute. These states we have identified with a neutron PDR. The
states in the region E*= 8-8.5 MeV in Z=50 and N=82 and E*= 9-9.5
MeV in N=50 nuclei carry a different signature, being compatible
with the low-energy part of the GDR. At E*= 9-20 MeV a strong,
isovector oscillation, corresponding to the excitation of the GDR is
obtained. An interesting observation is the most exotic $^{100}$Sn
nucleus, where at E*=8.3 MeV a state with a proton structure is
found. The analysis on dipole transition densities for different
excitation energy regions in $^{100}$Sn is presented in
Fig.\ref{fig5}, illustrating the proton surface oscillations at
E*$\leq$8.3 MeV. This mode could indicate a proton PDR. The
dependence of the calculated total PDR strength on the mass number
in N=50,82 and $^{100-132}$Sn is compared to the relative difference
between the neutron and proton rms radii
\begin{equation}\label{eq:skin}
\delta r=\sqrt{<r^2_n>}-\sqrt{<r^2_p>}
\end{equation}
in the right hand side part of Fig.2, Fig.3 and Fig.4, respectively.
In the case of N=50 and N=82 isotones we keep the neutron number
fixed and change the proton number only. This affects the thickness
of the neutron skin (see Fig.\ref{fig1} left) as well and
respectively the total PDR strength (Fig.\ref{fig2} right) and
(Fig.\ref{fig3} right) decreases with increasing proton number. The
results obtained for $^{100-132}$Sn nuclei, where the neutron number
increases from N=50 to N=82 are in agreement with these obtained for
N=50,82 isotones considered above. Accordingly, the total PDR
strength increases (Fig.\ref{fig4} right), when $\delta$r increases
and correspondingly the neutron or proton skin thicknesses increase
(Fig.\ref{fig1} right).


\section{Conclusions}

In the isotones with N=50, N=82 and the Z=50 isotopes low-energy
dipole states, identified with PDR were obtained. A close connection
between the total PDR strengths and the neutron skin thickness
defined by the relative difference of neutron and proton rms radii
was found. These observations agree very well with our previous
results for the $Z=50$ isotopes and the $N=82$ isotones. In the most
exotic nuclei $^{100-104}$ Sn lowest dipole states of almost pure
proton structure are identified. They are related to oscillations of
weakly bound protons, indicating a proton-driven PDR. The
interesting point is, that these states are predicted to exist in
heavy nuclei with N slightly larger or equal to Z. We suggest, that
the effect is due to Coulomb repulsion, that pushes the weakly bound
protons orbitals into the nuclear surface. The results for Sn
isotopes and N=82 nuclei are in a good  agreement with available
data \cite{Volz06,Gov,Ozel}.

At present, extended investigations on the fragmentation pattern of
the low-energy dipole excitations are in progress. The QPM
calculations will be performed in considerably larger configuration
spaces and using microscopically derived interactions, thus enabling
a detailed description of data on the dipole response of stable and
exotic nuclei to be expected for the near future from ELBE and the
experiments planned at GSI and for FAIR.

\end{document}